\documentclass[twocolumn]{aa}
\usepackage{graphicx}
\begin{document}

\title{Non-LTE Abundances of Magnesium, Aluminum and Sulfur in OB Stars
Near the Solar Circle}

\author{S. Daflon \inst{1}, K. Cunha \inst{1}, V. V. Smith \inst{2}
          \and K. Butler\inst{3}}

\offprints{S. Daflon}

\institute{Observat\'orio Nacional, Rua General Jos\'e Cristino 77 
	   CEP 20921-400, Rio de Janeiro,  BRAZIL\\
           \email{daflon@on.br, katia@on.br}
         \and
	   Department of Physics, University of Texas at El Paso, 
	   El Paso, TX 79968-0515 USA \\
	   \email{verne@barium.physics.utep.edu} 
	 \and
           Institut f\"ur Astronomie und Astrophysik der Universit\"at 
	   M\"unchen, Scheinerstrasse 1, D-81679 M\"unchen, GERMANY \\  
           \email{butler@usm.uni-muenchen.de}
	   }

\date{Received ; accepted }

\abstract{
Non-LTE abundances of magnesium, aluminum and sulfur are derived
for a sample of 23 low-$v \sin i$ stars belonging to six northern 
OB  associations of the Galactic disk within 1 kpc of the 
Sun. The abundances are obtained from the fitting of synthetic 
line profiles to high resolution spectra. A comparison of our 
results with HII region abundances indicates good  agreement 
for sulfur while the cepheid abundances are higher. The derived 
abundances of Mg show good overlap with the cepheid results. 
The aluminum abundances for OB stars are significantly below
the cepheid values. But, the OB star results show a dependence 
with effective temperature and need further investigation. 
The high Al abundances in the cepheids could be the result of 
mixing.  A discussion of the oxygen abundance in objects near 
the solar circle suggests that the current mean galactic oxygen 
abundance in this region is 8.6-8.7 and in agreement with the 
recently revised oxygen abundance in the solar photosphere. 
Meaningful comparisons of  the absolute S, Al and Mg
abundances in OB stars with the Sun must await a reinvestigation
of these elements
with 3D hydrodynamical model atmospheres for the Sun.
No abundance gradients are found within the limited range in 
galactocentric distances in the present study. Such variations 
would be expected only if there were large metallicity gradients 
in the disk.
}

\titlerunning{Non-LTE Mg, Al, and S}
\authorrunning{Daflon et al.}
\maketitle

\section{Introduction}

The massive OB stars, as well as other young objects such as 
cepheids or H~II regions, are used commonly as tracers of the 
current chemical composition in the Galactic disk.  Abundance 
analyses of H~II regions are restricted to a handful of elements, 
those being most typically He, N, O, S, Ne, and Ar.  In the OB 
stars, besides N, O, and S, the additional elements C, Mg, Al, 
Si, and Fe can be analyzed using both LTE and non-LTE techniques.
Cepheids are the  more evolved, cooler descendents of a subset of 
the OB stars  (with masses of $\sim$ 5-10M$_{\odot}$) and a rather 
large number  of elements ($\sim$ 25) can be detected in their 
spectra.  As all three of these types of objects are young ($\le$ 
few $\times$ 10$^{7}$ yr), it is reasonable to expect that examples 
of them inhabiting the same region of the Galaxy should contain 
approximately the same mixture of chemical elements.  This 
approximation should be as good as allowed by small scale chemical 
inhomogeneities that can be produced on timescales of the lifetimes 
of large, starforming regions, i.e. a few times 10$^{7}$ yr.

Care must be taken in comparing abundances between H~II regions, 
OB stars, and cepheids, however. For example, internal stellar 
mixing may affect both OB stars and cepheids to varying degrees.  
Some evidence of the mixing of material exposed to the CN-cycle 
has been uncovered in certain OB stars, even near the main sequence 
(Gies \& Lambert \cite{gl92}); this mixing may be driven primarily 
by rotation (Heger \& Langer \cite{hel00}).  In this case, nitrogen 
abundances are measurably enhanced (by $\sim$ +0.3 dex)  with, 
perhaps, a marginal decrease in the carbon abundance 
(by $\sim$ $-$0.1 dex), while oxygen remains untouched.  In the 
more evolved cepheids, even deeper mixing may have occurred that 
involves the full CNO cycles, such that some oxygen depletion might 
be detectable (as well as larger N enhancements and C depletions).  
There is also the possibility that surface abundances of sodium, 
magnesium, and aluminum have been altered by the mixing of material 
exposed to the Ne-Na and Mg-Al cycles.  In the H~II regions, 
uncertainties or systematics may arise from temperature fluctuations 
in the gas, unknown radiation environments, or possible depletions 
of some elements out of the gas phase and onto solid particles. 

By concentrating on a set of OB stars, in comparison to sample 
H~II regions and cepheids, all near the solar circle, abundance 
trends found in these various objects can be intercompared 
and ultimately be checked against their corresponding solar values.  
Such checks can help uncover possible inconsistencies and provide 
stronger constraints on abundance gradients derived from different 
types of objects.  That is the aim of this paper, which is the 
fourth paper in a series whose ultimate goal is to use a large 
sample of OB stars to trace abundance gradients in the Galactic 
disk.  Paper I (Daflon, Cunha \& Becker \cite{pap1}) presented the 
first results from this survey  and concentrated on 8 sharp lined 
star ($v \sin i\le$ 60 km s$^{-1}$) members of the Cep~OB2 
association: LTE abundances of C, N, O, Si, and Fe were derived, 
as well as non-LTE abundances of C, N, O, and Si.  Daflon et al. 
(\cite{pap2} $-$ Paper II) added analyses of 15 members of five 
additional northern OB associations (Cep~OB3, Cyg~OB2, Cyg~OB7, 
Lac~OB1, and Vul~OB1).  The same sets of elements were analyzed in 
non-LTE as in Paper I, but additional LTE results were presented 
for Mg, Al, and S.  In Paper III  (Daflon et al. \cite{pap3}), 
the analysis was expanded to include some of the more rapidly 
rotating stars ($v \sin i$= 60-150 km s$^{-1}$) in the northern 
sample.  In addition, the atomic analysis included non-LTE 
calculations for Mg and Al. In this paper, the non-LTE calculations 
for Mg and Al are applied to the remaining northern stars, and 
sulfur is now added as an element that can be studied in non-LTE. 

\section{Observations and Stellar Parameters}

The observational data consist of high resolution (R$\sim$60\,000), 
high signal-to-noise spectra of 23 main sequence late-O/early-B 
stars belonging to the OB associations of Cep~OB2, Cep~OB3, 
Cyg~OB3, Cyg~OB7, Vul~OB1 and Lac~OB1. The spectra were obtained 
at the McDonald Observatory, University of Texas, Austin, with the 
2.1m telescope plus the Sandiford Echelle Spectrograph. A set of 
lower resolution (R=12\,000) spectra of these targets were obtained 
with the 2.7m  telescope plus a Coud\'e Spectrograph in the H$\gamma$ 
region. More details about the observations and data reduction 
are found in Papers I and II. Also in these studies are derived the
stellar parameters and microturbulences for the 23 target stars.  

The effective temperatures and surface gravities have been derived 
from a photometric calibration for the reddening-free index Q coupled
to the fitting of the broadened wings of H$\gamma$ profile. In our 
earlier series of papers  (Papers I and II) it has been argued that 
this method results in uncertainties of $\sim$4\% in $T_{eff}$ and
$\pm$0.10  dex in $\log g$. The uncertainties in gravity can be also 
viewed by noting the positions of stars in a $\log T_{eff} - \log g$ 
diagram in comparison to model tracks. For stars of these effective 
temperatures the $\log g$ of the ZAMS is about 4.2. An average of all 
surface gravities of stars here that are not clearly evolved (i.e., 
$\log g \le$ 4.0) finds $\log g=4.25 \pm 0.17$; close to what is 
expected for stars near the main sequence with a scatter larger than 
what we have estimated as the expected uncertainty. The microturbulent 
velocities were obtained from the requirement that the non-LTE  O~II 
abundances were independent of the line-strength. The sample stars, 
their corresponding associations and spectral types, adopted effective 
temperatures, surface gravities and $\xi$-value are gathered in 
Table~\ref{stepar}. We note that the surface gravity for the star 
HD~214167 has been revised. 

\begin{table}
\caption[]{Stellar Parameters}
\label{stepar} 
$$
\begin{array}{cccccc}
\hline
\noalign{\smallskip}
\mathrm{Association} & \mathrm{Star} & \mathrm{MK} & {T_{eff}\mathrm(K)} 
& \log g & \xi \mathrm{(km s^{-1})} \\
\noalign{\smallskip}
\hline
\noalign{\smallskip}
\mathrm{Cep OB2}& \mathrm{HD} 205794 & \mathrm{B0.5V} & 26890 &4.21 &  8.0 \\
       & \mathrm{HD} 206183          & \mathrm{O9.5V} & 33310 &4.52 &  5.0 \\
       & \mathrm{HD} 206267          & \mathrm{B0V}   & 26100 &4.21 &  5.0 \\
       & \mathrm{HD} 206327          & \mathrm{B2V}   & 21900 &3.99 &  8.0 \\
       & \mathrm{HD} 207538          & \mathrm{O9V}   & 32190 &4.32 & 10.0 \\
       & \mathrm{HD} 239724         & \mathrm{B1V} & 24790 &3.83 & 12.0 \\
       & \mathrm{HD} 239742         & \mathrm{B2V} & 22470 &4.07 &  5.0 \\
       & \mathrm{HD} 239743         & \mathrm{B2V} & 21580 &3.99 &  8.0 \\
\mathrm{Vul OB1}& \mathrm{BD}+24^{\circ} 3880 & \mathrm{B0.5V} & 30410 &4.57 &  8.0 \\
       & \mathrm{HD} 344783         & \mathrm{B0IV} & 31010 &4.26 &  9.0 \\
\mathrm{Cyg OB3}& \mathrm{HD} 227460  & \mathrm{B1IV} & 27060 &4.34 &  8.0 \\
       & \mathrm{HD} 227586         & \mathrm{B0.5V} & 27830 &4.15 &  8.0 \\
       & \mathrm{HD} 227757         & \mathrm{O9.5V} & 32480 &4.22 &  8.0 \\
\mathrm{Cyg OB7}& \mathrm{HD} 197512   & \mathrm{B1V} & 23570 &4.02 &  6.0 \\
       & \mathrm{HD} 202253         & \mathrm{B1.5IV} & 22750 &3.95 &  6.0 \\
\mathrm{Lac OB1}& \mathrm{HD} 214167  &  \mathrm{B1V} & 26720 & 4.54 &  6.0 \\
       & \mathrm{HD} 214680        & \mathrm{O9V}  & 33690 &4.27 & 11.0 \\
       & \mathrm{HD} 216916       & \mathrm{B2IV}  & 23520 &4.00 &  6.0 \\
       & \mathrm{HD} 217227       & \mathrm{B2V } & 19000 &4.20 &  7.0 \\
       & \mathrm{HD} 217811       & \mathrm{B2V}  & 19070 &3.92 &  5.0 \\
\mathrm{Cep OB3}&\mathrm{BD}+62^{\circ} 2125 &\mathrm{B1.5V} & 23480 &4.05 & 10.0 \\
       & \mathrm{HD} 217657         & \mathrm{B0.5V} & 27950 &4.38 &  8.0 \\
       & \mathrm{HD} 218342       & \mathrm{B0IV}  & 30020 &4.20 &  9.0 \\
\noalign{\smallskip}
\hline
\end{array}
$$
\end{table}

\section{Non-LTE Abundances}

Non-LTE synthetic spectra were calculated for the few transitions 
of Mg~II, Al~III and S~III that are available in the spectra
of early-type stars. Although relatively free of blends, linelists
for each spectral region containing the lines of interest were 
constructed  within the interval around $\pm$3\AA. These transitions 
are listed in Table~\ref{llist} together with their wavelengths, 
designations, excitation potentials as well as adopted gf-values. 

\begin{table}
\caption[]{Linelists}
\label{llist} 
$$
\begin{array}{ccccr}
\hline
\noalign{\smallskip}
\mathrm{Wavelength\,\, interval} & \lambda(\AA) & \mathrm{Species} & \chi(eV) &
\log (gf) \\
\noalign{\smallskip}
\hline
\noalign{\smallskip}
4359-4367 & 4361.53 & \mathrm{S\,III} & 18.24 & -0.75 \\
\mathrm{S\,III}    & 4364.75 & \mathrm{S\,III} & 18.32 & -0.85 \\
          & 4366.91 & \mathrm{O\,II}  & 23.01 & -0.24 \\
\hline
4476-4484 & 4479.88 & \mathrm{Al\,III} & 20.78 & 0.90* \\
\mathrm{Mg\,II, Al\,III} & 4479.97  & \mathrm{Al\,III} & 20.78 & 1.02* \\
          & 4481.13 & \mathrm{Mg\,II} & 8.86 & 0.74 \\
          & 4481.15 & \mathrm{Mg\,II} & 8.86 & -0.56 \\
          & 4481.33 & \mathrm{Mg\,II} & 8.86 & 0.59 \\
\hline
4510-4515 & 4510.88 & \mathrm{N\,III}  & 35.67 & -0.45 \\
\mathrm{Al\,III}    & 4512.56 & \mathrm{Al\,III} & 17.81 & 0.40 \\
          & 4514.85 & \mathrm{N\,III}  & 35.70 & 0.23 \\
\hline
4526-4531 & 4527.86 & \mathrm{N\,III}  & 38.49 & -0.24 \\
\mathrm{Al\,III}    & 4529.19 & \mathrm{Al\,III} & 17.81 & 0.66 \\
          & 4530.41 & \mathrm{N\,II}   & 23.49 & 0.67 \\
\noalign{\smallskip}
\hline
\end{array}
$$
\begin{list}{}{}
\item[$^{*}$] $\log(gf)$ from Kurucz web site; all the others
gf-values are from TOPbase.
\end{list}
\end{table}

The abundance analyses in this study are based on the fully-blanketed 
and pla\-ne-parallel LTE model atmospheres calculated with the ATLAS9 
code (Kurucz \cite{kur92}) for a  constant microturbulent velocity of 
$\xi=2{\rm km s^{-1}}$ and solar composition. Departures from LTE 
were considered in the line formation calculations with the  newest 
version of the program  DETAIL (Butler \cite{det}).  The adopted model 
atoms are described in  Przybilla  et al. (\cite{nob01} - Mg~II), 
Dufton et al. (\cite{duf86} - Al~III) and  Vrancken  et al.  
(\cite{vra96} -  S~III). We note that so far, all abundance papers 
including non-LTE line formation of Al published in the literature 
(Vrancken et al. \cite{vra97}; Vrancken et al. \cite{vra00}; 
Gummersbach et al. \cite{gum98}) were based on the same model atom 
adopted in this study, and in a sense are not completely independent. 
Published non-LTE  abundance analyses of magnesium (Vrancken et al. 
\cite{vra97}; Vrancken et al. \cite{vra00}; Gummersbach et al. 
\cite{gum98}) are based on the model atom of Mihalas (\cite{mih72}).  
All sulfur abundances in the literature are derived from LTE analysis, 
except for Vrancken et al. (\cite{vra96}), who analyzed three B stars 
as a test for their sulfur model atom. The present study is the first 
systematic non-LTE analysis  of sulfur abundances in OB stars. 

A brief description of the adopted model atoms  follows.
Przybilla et al. (\cite{nob01}) constructed an extensive model atom 
for Mg~I/Mg~II based on recent atomic data. This model is roughly 
complete for levels up to n=9 for Mg~I and n=10 for Mg~II, whereas  
Mg~III, that does not have a significant population in  the range of 
effective temperatures considered here, is represented only by its 
ground state.  The sulfur model atom (Vrancken et al. \cite{vra96}) 
treats S~II/S~III simultaneously considering  81 levels of S~II and 
21 levels of S~III. The model atom  also includes the three lowest 
levels of S~I and the two lowest levels of S~IV, together with 
the ground state of S~V that are important for the hotter stars.
Vrancken et al. (\cite{vra96}) tested their model atom  for three 
B2~III-V stars and concluded that S~III lines yield more reliable 
abundances (than S~II lines) for the temperature range they consider. 
The adopted Al~III model atom of Dufton et al. (\cite{duf86}) is 
less complete than the magnesium and sulfur ones. It consists of  12  
states of Al~III plus the ground state of Al~IV, as the populations of 
Al~II and Al~V are not significant in the temperature range between 
20000 to 35000K.

\begin{figure}
   \centering
   \includegraphics[angle=-90,width=\columnwidth]{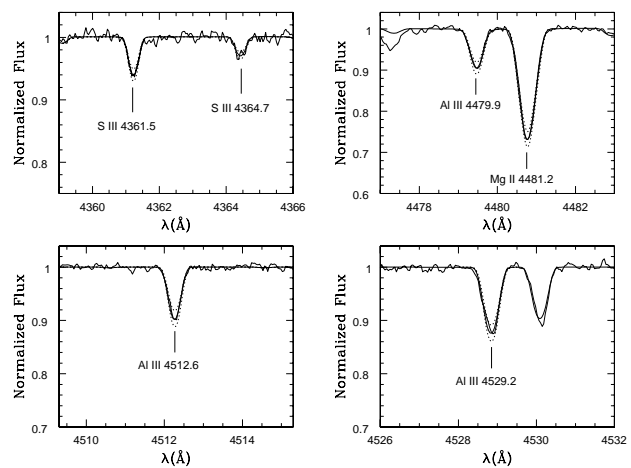}
      \caption{Some examples of line synthetic profiles fitted 
to the observed spectra of the star HD~197512 for 
$\xi=6 {\rm km s^{-1}}$. The solid lines represent the best 
fit obtained for each observed line profile, identified by its 
wavelength in \AA, whereas the dotted lines depict the theoretical 
profiles computed for abundances sligthly lower and higher 
($\pm$0.1 dex) than the adopted value for each element. }
         \label{fitsample}
   \end{figure}

The synthetic line profiles were calculated with the SURFACE code
(Butler \cite{srf}), assuming Voigt profile functions. These profiles 
were then broadened by means of convolution with the rotational 
profile, including $v \sin i $ and limb darkening, and instrumental 
profile. The  microturbulence for each target star was adopted from 
O II lines, derived in previous studies, while the line abundances 
and $v \sin i$'s were allowed to vary. The best fit was chosen from 
the $\chi^2$-minimization of the differences between theoretical and 
observed profiles. The final non-LTE abundances and $v \sin i$'s are 
listed in Table~\ref{nonab}. The abundances are represented by the 
average of the individual line abundances and the respective 
dispersions (and number [n] of fitted lines), whenever this is the 
case. A sample of profile fitting for all the spectral regions is 
shown in the panels of Figure~\ref{fitsample}, for the star HD~197512. 

\begin{table*}
\caption[]{Non-LTE Mg, Al, and S Abundances}
\label{nonab} 
$$
\begin{array}{cccccc}
\hline
\noalign{\smallskip}
\mathrm{Association} & \mathrm{Star} & v \sin i(\mathrm{km s}^{-1}) & \log
\epsilon(Mg) [n] & \log \epsilon(Al)  [n]  & \log \epsilon(S) [n] \\
\noalign{\smallskip}
\hline
\noalign{\smallskip}
\mathrm{Cep OB2}& \mathrm{HD} 205794         & 13.2\pm1.2 & 7.33 [1] &
5.81\pm0.16 [2] & 7.34  [1]         \\
       & \mathrm{HD} 206183         & 19.2\pm1.9 & 7.36 [1] & 6.06\pm0.19 [3] &
7.05  [1]         \\
       & \mathrm{HD} 206267         & 47.0\pm3.0 & 7.21 [1] & 5.74\pm0.15 [3] &
7.28  [1]         \\
       & \mathrm{HD} 206327         & 20.6\pm2.1 & 7.27 [1] & 5.98\pm0.05 [3] &
7.17\pm0.05 [2] \\
       & \mathrm{HD} 207538         & 42.8\pm2.0 & 7.42 [1] & 6.15\pm0.39 [2] &
...          \\
       & \mathrm{HD} 239724         & 38.0\pm1.0 & 7.48 [1] & 5.79\pm0.24 [3] &
7.16\pm0.16 [2] \\
       & \mathrm{HD} 239742         &  9.5\pm1.2 & 7.37 [1] & 5.96\pm0.07 [3] &
7.19  [1]         \\
       & \mathrm{HD} 239743         & 23.2\pm1.3 & 7.33 [1] & 6.01\pm0.03 [3] &
...           \\
\mathrm{Vul OB1}& \mathrm{BD}+24^{\circ} 3880 & 12.9\pm3.3 & 7.59 [1] &
6.24\pm0.19 [3] & 7.28\pm0.08 [2] \\
       & \mathrm{HD} 344783         & 29.2\pm4.3 & 7.31 [1] & 5.95\pm0.42 [2] &
7.13\pm0.03 [2] \\
\mathrm{Cyg OB3}& \mathrm{HD} 227460         & 14.2\pm2.8 & 7.46 [1] &
6.11\pm0.17 [3] & 7.22  [1]         \\
       & \mathrm{HD} 227586         & 19.7\pm3.2 & 7.21 [1] & 5.79\pm0.11 [2] &
7.21  [1]         \\
       & \mathrm{HD} 227757         & 29.6\pm3.2 & 7.39 [1] & 6.29\pm0.23 [3] &
7.34  [1]         \\
\mathrm{Cyg OB7}& \mathrm{HD} 197512         & 16.5\pm1.2 & 7.38 [1] &
5.98\pm0.07 [3] & 7.11\pm0.19 [2] \\
       & \mathrm{HD} 202253         & 48           & ...  & 5.96 [1]          &
...           \\
\mathrm{Lac OB1}& \mathrm{HD} 214167         & 21.6\pm4.4 & 7.57 [1] &
6.01\pm0.09 [3] & 7.09  [1]         \\
       & \mathrm{HD} 214680         & 27.2\pm4.5 & 7.47 [1] & 6.28\pm0.31 [3] &
...           \\
       & \mathrm{HD} 216916         & 14.3\pm3.3 & 7.52 [1] & 6.12\pm0.16 [3] &
7.29  [1]         \\
       & \mathrm{HD} 217227         & 18.5\pm2.4 & 7.52 [1] & 6.18\pm0.08 [3] &
7.28  [1]         \\
       & \mathrm{HD} 217811         & 9.4\pm0.6  & 7.46 [1] & 6.05\pm0.04 [3] &
7.11\pm0.18 [2] \\
\mathrm{Cep OB3}&\mathrm{BD}+62^{\circ} 2125 & 39.6\pm3.2 & 7.03 [1] &
5.84\pm0.01 [2] & 7.20\pm0.21 [2] \\
       & \mathrm{HD} 217657         & 17.4\pm2.4 & 7.12 [1] & 5.71\pm0.11 [2] &
7.17  [1]         \\
       & \mathrm{HD} 218342         & 35.2\pm4.8 & 7.44 [1] & 5.99\pm0.01 [2] &
7.06\pm0.08 [2] \\
\noalign{\smallskip}
\hline
\end{array}
$$
\end{table*}

In general, the published  non-LTE Mg, Al and S abundances  are 
consistent with our results. However, aluminum abundances deserve 
special attention. As a test, we recalculated the non-LTE abundances 
of  star 201 in NGC 2244, analyzed by Vrancken et al. (\cite{vra97}), 
using their listed  equivalent widths and, as expected, we reproduced 
their derived abundance within the errors. We also compared our 
non-LTE  Al abundances  with those interpolated in the grid of 
theoretical equivalent widths of Dufton et al. (\cite{duf86}). The 
comparison  for the  model atmospheres  with lower $T_{eff}$ showed 
that our abundances derived directly from  synthesis agree within the 
uncertainties with the abundances interpolated within their grid. 
For the hottest models, however, Dufton's grid yields aluminum 
abundances  much lower than 6.0.  The  difference between our 
synthetic profiles and Dufton's  grid resides basically in the model 
atmospheres, as they used non-LTE non-blanketed model atmospheres 
and we use fully blanketed LTE models from Kurucz (\cite{kur92}).

Errors in the determination of the stellar parameters, microturbulence, 
$v \sin i$ and the placement of the continuum are the main sources of 
uncertainties in a chemical analysis based on the fitting of line 
profiles. In paper III we discussed the uncertainties assigned for 
each of these parameters and the subsequent abundance error arising 
from them. Accordingly, our total errors expected for the derived 
magnesium abundances are of the order of 0.2 dex, being larger for 
the coolest stars (0.3 dex for $T_{eff}\sim$19\,000K) and these are 
dominated by the uncertainty in the microturbulent velocity (as the 
abundance analysis is based on {\it one} intermediate to strong line) 
and  effective temperature. The uncertainties in sulfur abundances 
arise chiefly from the errors in $T_{eff}$ and are estimated to be 
of the order of $\sim$0.15 dex. The profile of the weak S~III line 
at 4364\AA \ was only fitted for the stars with lowest $v \sin i$ 
and highest signal-to-noise spectra. For this reason, the sulfur 
abundances of some stars in our sample are based only on the S~III 
line at 4361\AA. Formally, aluminum presents smaller errors, around 
0.10 dex. However, the abundances derived from  the lines 
$\lambda\lambda$ 4512 and 4529\AA, that belong to the same multiplet, 
show some dependence with effective temperature and introduce a larger 
discrepancy in the average abundances of the hottest stars.

\section{Discussion}

OB stars, as well as other young objects, such as H~II regions,
track the chemical composition of the Galactic disk.  With this 
paper, we now have a uniform set of non-LTE abundances for C, N, 
O, Mg, Al, Si, and S in 35 stars of the northern sample.  All 
six OB associations that are represented in this sample lie within 
1 kpc of the solar circle  (where we take R$_\odot$=7.9 kpc,   
McNamara et al. (\cite{mac00})) and, as such, will have abundances 
that are, in principle, not affected significantly by possibly modest 
galactic abundance gradients. The abundances of certain elements 
like sulfur, magnesium and aluminum, as well as oxygen can be 
compared to recent results derived from cepheids and H~II regions 
also lying within the same galactocentric distance interval.  

\begin{figure}
   \centering
   \includegraphics[width=\columnwidth]{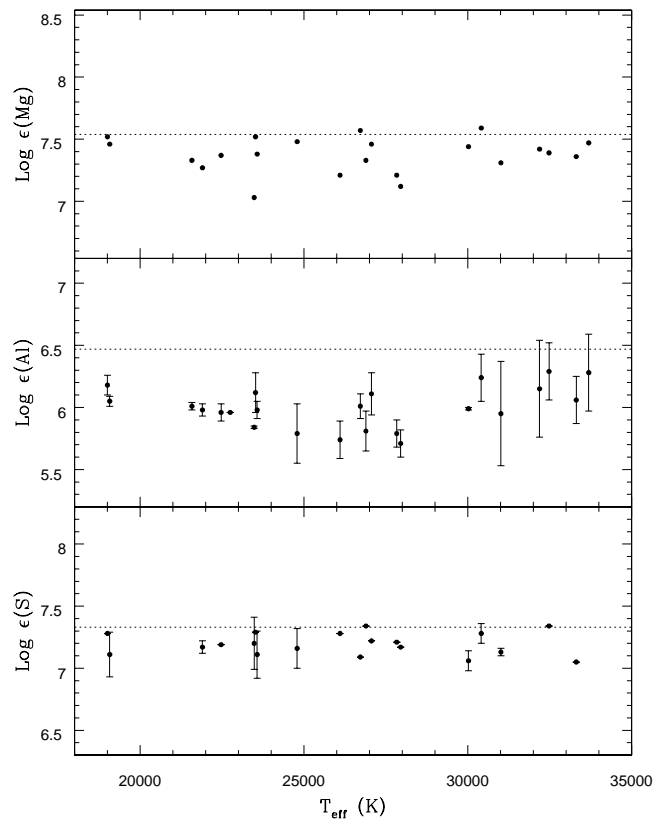}
      \caption{ Elemental abundances of magnesium, aluminum and sulfur as 
a function of effective temperature. 
The abundances are represented by the average of individual lines while 
errorbars represent the respective standard deviations.  
The dotted lines represent the
solar abundances listed by  Holweger (\cite{hol01}, Mg: 7.538$\pm$0.060) and
Grevesse \& Sauval (\cite{ges98}, Al: 6.47$\pm$0.07, S: 7.33$\pm$0.11). } 
         \label{tefab}
   \end{figure}

The stars in this sample are all sharp-lined stars, with $v \sin i$
$\le$ 48 km s$^{-1}$.  The mean magnesium abundance is 
A(Mg)= 7.37$\pm$0.14. The magnesium abundances are quite independent
of effective temperature, as shown in the top panel of 
Figure~\ref{tefab}. Aluminum shows significantly lower abundances 
in the OB stars than the Sun, as displayed in the middle panel of 
Figure~\ref{tefab}. Such a large difference may reside partially 
in uncertainties in the non-LTE calculations for Al~III.   Some 
evidence for this possibility may appear in the derived aluminum 
abundances showing a small offset between the hotter and cooler 
stars.  There is a noticeable displacement between stars with 
$T_{eff}\ge$30000K and those with lower effective temperatures, 
with the hotter sample having a mean aluminum abundance of 
6.13$\pm$0.14 and the cooler having 5.94$\pm$0.14. For the S~III 
non-LTE abundances, there is no trend with $T_{eff}$  and the 
scatter is small: the mean  A(S)= 7.19$\pm$0.09 is 0.14 dex  below 
the solar abundance of 7.33$\pm$0.11 (Grevesse \& Sauval 1998).  

Taken together, these new results for S, Mg and Al, could in principle 
suggest that the OB stars, on average, are slightly underabundant when 
compared to the Sun. However, these differences  could be argued to be 
at the level of systematics. In fact, the recent results from 
3D-hydrodynamical model atmosphere calculations for the elements 
C, N and O in the solar photosphere have lowered their abundances 
significantly. The photospheric carbon, nitrogen and oxygen 
abundances,  for instance, have been recently  revised to 8.40, 7.81, 
and 8.65, respectively (Asplund \cite{asp02}), and are now in better
agreement with the results for OB stars. 
In addition, 
the OB results rely on 1D-models while the recent solar results are 
from 3D hydrodynamical models. Although it is expected that the  
radiative  OB atmospheres will not have temperature inhomogeneities 
that are present in the solar atmosphere (caused by convection) it is 
possible that some other effects like the high microturbulent velocities 
required for the OB stars indicates that 1D models are not a complete 
description of atmospheres of hot stars. 

\begin{figure}
   \centering
   \includegraphics[angle=-90,width=\columnwidth]{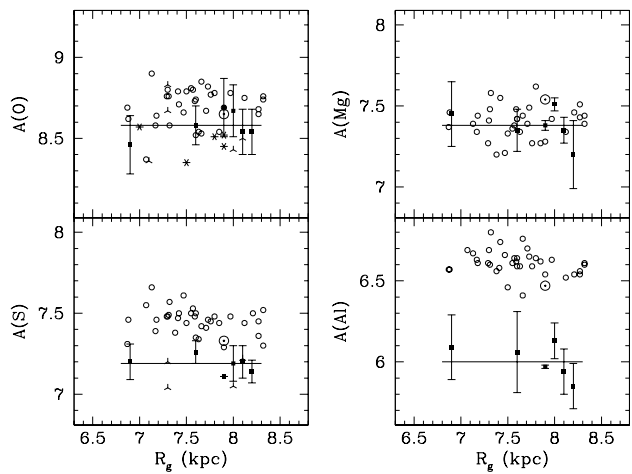}
      \caption{ A comparison between abundances for OB stars, 
cepheids, and H II regions, for the elements O, Mg, Al, and S. 
The abundances are  plotted as a function of  the Galactocentric 
distance, within 1 kpc from the Sun, represented at 
R$_\odot$= 7.9 kpc. The mean abundances of each OB 
association are plotted as filled squares, with the errorbars 
representing the standard deviation. The straight lines depict 
the average abundance for the six OB  associations for each element.
Abundances for cepheids (Andrievsky et al. \cite{and02}) are  shown as open 
circles. The abundances for H~II regions are represented by six-pointed 
stars (Deharveng et al. \cite{deh00}), and three-pointed stars (Simpson et al. 
1995; Afflerbach et al. \cite{aff97}).}
         \label{hiireg}
   \end{figure}

Abundances for the different types of young objects, OB stars,
cepheids, and H~II regions, are compared in Figure~\ref{hiireg} 
for the elements Mg, Al, and S, as well as O (with oxygen taken 
from our previous papers).  We restrict our discussion to 
galactocentric distances that overlap those values found for the 
six OB associations represented here, where R$_{\rm g}$= 6.9 to 
8.2 kpc.  Oxygen is added to the discussion in Figure~\ref{hiireg} 
as it represents the most abundant element after H and He, and is
the element comprising the largest fraction of a star's abundance 
of heavy elements.  In Figure~\ref{hiireg}, the mean abundances
of each OB association are plotted as filled squares, with the
errorbars representing the standard deviation found within the
abundances of the association members (in the case where only
2 members were represented, the errorbars are the average
differences from the mean).  Over the limited range of distance
sampled by this particular set of stars, no significant abundance 
gradient is apparent and the horizontal solid lines are the mean 
abundances of the set of OB associations.  The solar symbol is 
plotted at R$_\odot$= 7.9 kpc.  The mean oxygen abundance (non-LTE 
calculations for O II from Papers I, II, and III) is A(O)= 8.58$\pm$0.09.
Again, as discussed above, in agreement with the revised solar value.

Other results plotted in Figure~\ref{hiireg} include the recent 
cepheid results from Andrievsky et al. (\cite{and02}), shown as 
open circles, and H~II regions from the optical emission lines 
analyzed by Deharveng et al. (\cite{deh00}), shown as six-pointed 
stars, and H~II region far-IR fine-structure lines by Simpson et al. 
(\cite{sim95}) and Afflerbach et al.  (\cite{aff97}), shown as 
three-pointed stars.  

Beginning with oxygen in Figure~\ref{hiireg}, with abundances shown 
for cepheids and H~II region optical and far-IR lines, there is 
considerable overlap among all sets of results.  The cepheid 
O-abundances tend to be slightly higher than the OB associations, 
but only by $\sim$0.1 dex: likely within possible systematic 
offsets.  The optical H~II region results from Deharveng et al. 
(\cite{deh00}) scatter almost perfectly within the OB association 
results, while the far-IR results exhibit a  larger scatter, but 
with an average O-abundance that agrees with the OB associations.  
Overall, the suggestion from all of the oxygen abundances is 
that the current mean Galactic abundance near the solar circle 
is about A(O)$\sim$ 8.6-8.7, with an intrinsic dispersion that 
is not yet well-defined.  The dispersion among the six OB 
associations is quite small (as well as the H~II region optical 
emission-line results).

The panel of Mg and Al abundances in Figure~\ref{hiireg} contains 
only  OB association and cepheid results.  In the case of Mg, the 
overlap between OB stars and cepheids is essentially perfect.  
The mean and standard deviations are A(Mg)= 7.37$\pm$0.14 for the 
OB associations and 7.36$\pm$0.10 for the cepheids.  This tidy 
agreement between cepheids and OB stars for Mg is not repeated 
for Al, where the cepheid Al abundances are measurably above solar, 
while the OB stars are well below solar.  As discussed previously, 
the OB star abundances are based upon non-LTE calculations using
Al~III lines and there is a noticeable offset in the abundances
derived for the hotter and cooler stars.  This may indicate
problems in the non-LTE calculations.  On the other hand, the
rather large Al abundances for the cepheids may not be intrinsic,
but could be the result of stellar mixing involving material
exposed to the Mg-Al cycle of H-burning.  An increase in aluminum
would come from a decrease in Mg, however, the observed offset
in the Al abundances seen in the cepheids would result in a
negligible decrease in Mg, due to the larger Mg abundance.  
For example, if the current Galactic Al abundance is about solar, 
then the cepheids have had their Al abundances enhanced by 
$\sim$ 0.2 dex.  If this enhancement came from Mg atoms, it 
would result in a decrease in the overall Mg abundance
by $\sim$ 0.02-0.03 dex, which is effectively unmeasurable. 
More work will be needed to decide the best value for the 
current Galactic aluminum abundance at the solar circle.

Finally, Figure~\ref{hiireg} also shows sulfur abundances derived 
from the OB stars, cepheids, and H~II regions (from the far-IR 
lines). Here, the OB associations and H~II regions again show a 
great degree of overlap. The cepheids exhibit systematically larger 
S abundances than the Sun, and these enhancements cannot be due to 
internal stellar mixing.  

\section{Conclusions}

We have presented non-LTE abundances of magnesium, aluminum and 
sulfur for 23 OB stars members of OB associations within 1 Kpc of 
the Sun. Magnesium abundances derived for cepheids in the 
literature agree well with the results for OB stars while no Mg 
abundances can be derived for H II regions.  On the other hand, 
sulfur results point to discrepencies between the cepheids on 
one side and the OB stars and H~II regions on the other, that are 
not easily resolved: the cepheid sulfur abundances are derived from 
S~I lines, while the OB stars use S~III lines, and the H~II region 
results come from far-IR [S~III] lines at 19 $\mu$m.  Aluminum 
abundances OB stars  must be interpreted with  caution: the relatively 
low Al abundances derived may reflect problems in the non-LTE 
calculations for this species and deserves additional analysis before 
further conclusions while Al in the cepheids could be enhanced due to 
internal mixing. 

\begin{acknowledgements}
S.D. acknowledges a CAPES fellowship and partial financial support 
from DAAD (Germany). KC thanks David Lambert for travel support 
for observing runs in 1992, 1993, and 1994. VVS acknowledges support 
by the National Science Foundation through grant AST99-87374.
 
\end{acknowledgements}

\end{document}